\documentclass{PoS}

\usepackage{graphicx}

\usepackage{xcolor}
\colorlet{alert}{red!60!black}
\colorlet{example}{green!60!black}
\colorlet{structure}{blue!60!black}

\newcommand{\note}[1]{}                    

\newcommand{\half}{\mbox{\small $\frac{1}{2}$}}          

\newcommand{\ind}[1]{\rm\scriptscriptstyle #1}

\def\lsim{\mathrel{\rlap{\lower4pt\hbox{\hskip1pt$\sim$}}
    \raise1pt\hbox{$<$}}}                
\def\gsim{\mathrel{\rlap{\lower4pt\hbox{\hskip1pt$\sim$}}
    \raise1pt\hbox{$>$}}}                

\title{
\vspace*{-0.75cm}
\begin{minipage}{\textwidth}
\begin{flushright}
\texttt{\footnotesize
PoS(LATTICE2018)119   \\
ADP-18-32/T1080       \\
DESY 18-220           \\
Liverpool LTH 1190    \\
}
\end{flushright}
\end{minipage}\\[15pt]
\vspace*{+0.75cm}
       The strange quark contribution to the spin of the nucleon}

\ShortTitle{The strange quark contribution to \ldots}

\author{\speaker{R.~Horsley}$^{\,a}$,
        Y.~Nakamura$^b$,
        H.~Perlt$^c$,
        D.~Pleiter$^{d}$, 
        P.~E.~L. Rakow$^e$,
        G.~Schierholz$^f$,
        A.~Schiller$^c$,
        H.~St\"uben$^g$,
        R.~D.~Young$^h$
        and J.~M. Zanotti$^h$ \\
        \llap{$^a$} School of Physics and Astronomy,
                    University of Edinburgh,
                    Edinburgh  EH9 3FD, UK \\
        \llap{$^b$} RIKEN Advanced Institute for Computational Science,
                    Kobe, Hyogo 650-0047, Japan \\
        \llap{$^c$} Institut f\"ur Theoretische Physik,
                    Universit\"at Leipzig, 04109 Leipzig, Germany \\
        \llap{$^d$} JSC, Forschungszentrum J\"ulich,
                    52425 J\"ulich, Germany, \\
        \llap{\phantom{$^d$}} 
                    Institut f\"ur Theoretische Physik,
                    Universit\"at Regensburg, 93040 Regensburg, Germany \\
        \llap{$^e$} Theoretical Physics Division,
                    Department of Mathematical Sciences,
                    University of Liverpool, \\
        \llap{\phantom{$^e$}} 
                    Liverpool L69 3BX, UK \\
        \llap{$^f$} Deutsches Elektronen-Synchrotron DESY,
                    22603 Hamburg, Germany \\
        \llap{$^g$} Universit\"at Hamburg, Regionales Rechenzentrum,
                    20146 Hamburg, Germany \\
        \llap{$^h$} CSSM, Department of Physics,
                    University of Adelaide, Adelaide SA 5005, Australia \\
        E-mail: \email{rhorsley@ph.ed.ac.uk} }

\author{QCDSF-UKQCD Collaborations}

\abstract{Quark line disconnected matrix elements of an operator, such
          as the axial current, are difficult to compute on the lattice.
          The standard method uses a stochastic estimator of the operator,
          which is generally very noisy. We discuss and develop further
          our alternative approach using the Feynman-Hellmann theorem
          which involves only evaluating two-point correlation functions.
          This is applied to computing the contribution of the quark spin
          to the nucleon and in particular for the strange quark.
          In this process we also pay particular attention to the development
          of an SU(3) flavour breaking expansion for singlet operators.}

\FullConference{The 36th Annual International Symposium on
                Lattice Field Theory - LATTICE2018\\
		22-28 July, 2018\\
		Michigan State University, East Lansing, Michigan, USA.}

\begin{document}


\section{Introduction/Approach}
\label{approach}


The proton consists of two valence up quarks and one down quark
together with a `sea' of quark anti-quark pairs and gluons.
How each constituent contributes to the total spin of the proton
has remained a mystery for many years. In particular the quark
contribution is much smaller than expected from the naive quark
model. We discuss here our lattice QCD determination of the quark
contribution, using a novel technique, based on a field theoretic
application of the Feynman-Hellmann theorem,
\cite{Chambers:2014qaa,Chambers:2015bka}.

There are two common spin decompositions or `schemes': 
Jaffe--Manohar (JM), \cite{Jaffe:1989jz}, and Ji, \cite{Ji:1996ek}.
They both have a common quark spin term, $\Delta\Sigma/2$ but other
pieces vary. In particular the JM approach has a gluon spin piece,
$\Delta G$, which can be measured in $pp$ machines, while the Ji
approach is more suitable for polarised DIS and DVCS processes
and also lattice QCD determinations.

The Ji gauge invariant decomposition of the proton spin derived from
the symmetric energy--momentum tensor is given by
\begin{eqnarray}
   {1 \over 2} = {1\over 2}\Delta\Sigma_p + \sum_q L_q + J_g \,,
\end{eqnarray}
where $L_q$ is the orbital angular momentum of valence quark $q$ and
$J_g$ is the gluon angular momentum. We shall not discuss these terms
further here. The total quark spin
$\Delta\Sigma_p = \Delta\Sigma^{\ind{con}}_p + \Delta\Sigma^{\ind{dis}}_p$ with
\begin{eqnarray}
   \Delta\Sigma^{\ind{con}}_p 
      = \Delta u^{\ind{con}}_p + \Delta d^{\ind{con}}_p \,, \qquad
   \Delta\Sigma^{\ind{dis}}_p 
      = \Delta u^{\ind{dis}}_p + \Delta d^{\ind{dis}}_p + \Delta s^{\ind{dis}}_p \,,
\end{eqnarray}
where $\Delta q^{\ind{con},\, \ind{dis}}_p$ are the quark line connected
and disconnected proton, $p$, matrix elements of the axial current
respectively. We shall discuss the disconnected matrix elements further
in the next section noting here that for the proton there is only
a disconnected piece for the strange quark, so $\Delta s_p^{\ind{dis}}
\equiv \Delta s_p$. Similar relations also hold for the other members
of the baryon, $B$, nucleon octet.

The `Spin crisis', discovered many years ago is that $\Delta \Sigma_p$ is
small and only around $\sim 35\%$ of total spin, whereas in the naive
quark model it would be expected that the valence quarks give the complete
contribution $\Delta\Sigma_p \sim 1$. Here we shall consider
$\Delta\Sigma^{\ind{dis}}_B$ and the $\Delta s^{\ind{dis}}_B$ pieces.


\section{Feynman--Hellmann applied to field theories}


If we modify the action by $S(\lambda) = S + \lambda O$, then it can be
shown that \cite{Chambers:2014qaa}
\begin{eqnarray}
    {\partial E_B(\lambda) \over \partial \lambda}
        = {1 \over 2E_B(\lambda)}
                  \left\langle B \left|
                     : \widehat{O} :
                       \right|B \right\rangle_\lambda \,,
\end{eqnarray}
(where $:\ldots:$ means that the vacuum term has been subtracted.)
Thus by suitably choosing $O$ and by identifying numerically
the gradient of $E_B(\lambda)$ at $\lambda = 0$ we can determine
the desired matrix element. The computation requires only
$2$-point correlation functions (rather than the more complicated
$3$-point functions).

\begin{figure}[!htb]
   \begin{center}
   \begin{minipage}{0.40\textwidth}
      \includegraphics[width=5.00cm]{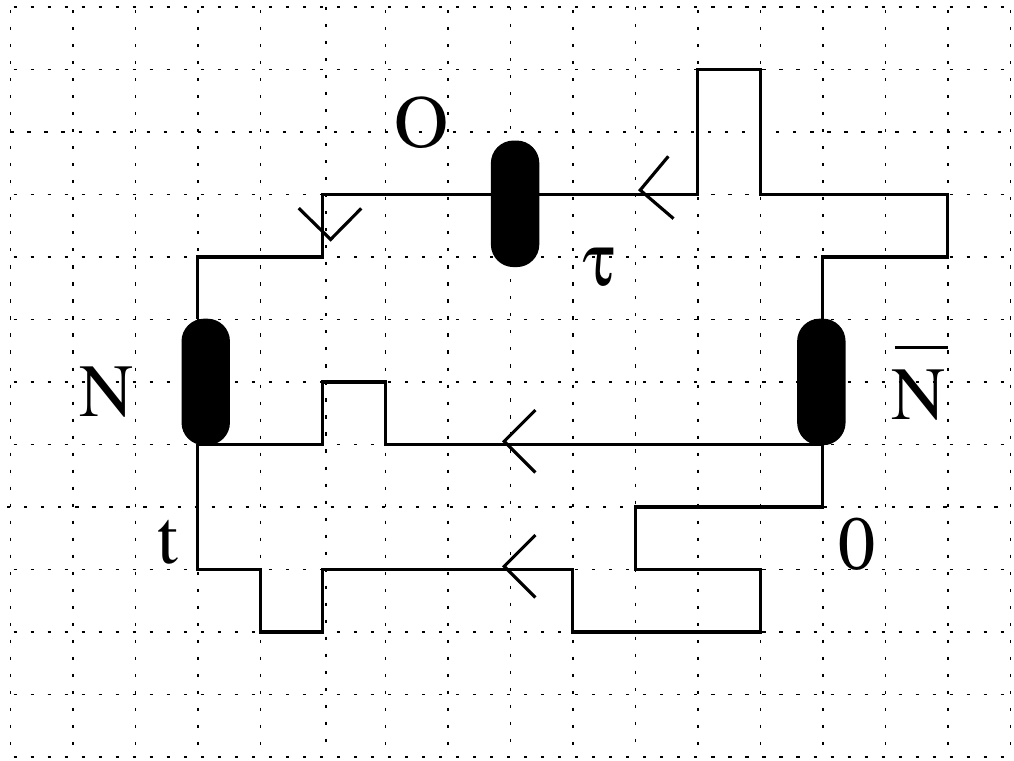}
   \end{minipage} \hspace*{0.10\textwidth}
   \begin{minipage}{0.40\textwidth}
      \includegraphics[width=5.00cm]{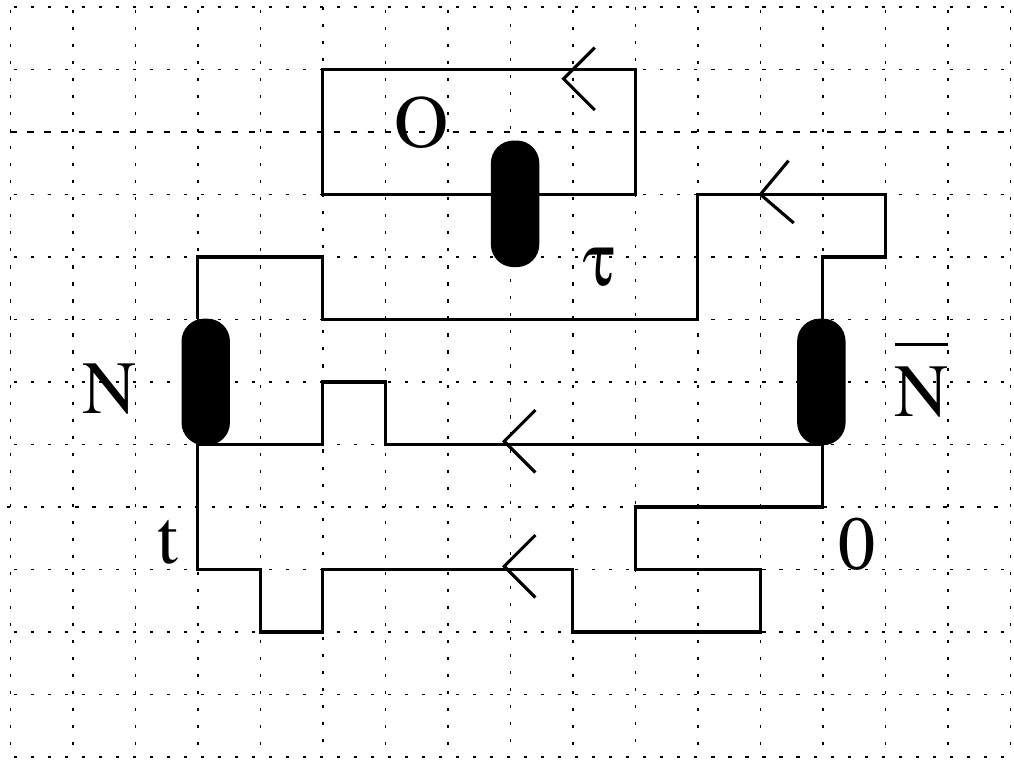}
   \end{minipage}
   \end{center}
   \caption{Left panel: Quark line connected $3$-point correlation
            functions;
            Right panel: Quark line disconnected $3$-point correlation
            functions.}
   \label{B_3pt}
\end{figure}

The modification location determines the contributions we access,
as indicated in Fig.~\ref{B_3pt}.
We can modify the Dirac fermion matrix before quark propagator inversion
\begin{eqnarray}
   D^{\prime\, -1} = \left[ D + \lambda O\right]^{-1}
   \quad \Rightarrow \quad
   \left. {\partial \over \partial \lambda} D^{\prime\,-1} \right|_{\lambda = 0}
                = D^{-1} O D^{-1} \,,
                                                         \nonumber
\end{eqnarray}
which inserts connected contributions on the quark line or we can
modify the field weighting during the HMC
\begin{eqnarray}
   \det D^\prime \, e^{-S_g} = \det [D + \lambda O] \, e^{-S_g} 
   \quad \Rightarrow \quad
   \left. {\partial \over \partial \lambda} \det D^\prime \right|_{\lambda = 0}
                           = \mbox{tr}(D^{-1}O) \, \det D \,,
\end{eqnarray}
which acesses disconnected contributions. (Or do both modifications and
obtain both connected and disconnected terms.) While the connected
piece is easy to implement, the disconnected piece requires the generation
of new configurations.

For a nucleon polarised in the $z$-direction we have
\begin{eqnarray}
   \langle B, \sigma | i\, \overline{q}\gamma_3\gamma_5 q 
                           | B, \sigma \rangle 
          = 2 M_B \sigma \Delta q\, \qquad \sigma = \pm \,,
\label{NME}          
\end{eqnarray}
which may be determined by applying the FH theorem to
\begin{eqnarray}
   C_\sigma(\lambda, t)
      \equiv (\Gamma_\sigma)_{\beta\alpha} 
             \langle B_\alpha(t)\overline{B}_\beta(0) \rangle_\lambda
       = A_B(\sigma\lambda) e^{-E_B(\sigma\lambda)} \,,
\end{eqnarray}
with corresponding projection operator
$\Gamma_\sigma = \half ( 1 + \gamma_4 )( 1 + i\sigma \gamma_3\gamma_5 )$.
As can be seen from eq.~(\ref{NME}) flipping the sign of $\lambda$
is equivalent to flipping the spin polarisation, so we can write the
amplitude and energy as a combined function of $\sigma\lambda$ .
For the connected contributions this is sufficient, but a further
complication arises for the disconnected terms, as for the generation
of configurations using HMC the fermion matrix in the action must be 
$\gamma_5$-hermitian for HMC, i.e.\ we now need
\begin{eqnarray}
   S = S_g + \sum_{q\,x} \lambda_q \overline{q}(x)\gamma_3\gamma_5 q(x) \,,
\end{eqnarray}
(rather than for the connected pieces,
$D^\prime = D + i \sum_{q\,x} \lambda_q \overline{q}(x)\gamma_3\gamma_5 q(x)$).
The correlation function thus develops imaginary parts in both
the amplitude
$A_B(\sigma\lambda) \to A_B(\sigma\lambda) e^{i\delta(\sigma\lambda)}$,
and energy
$E_B(\sigma\lambda) \to E_B(\sigma\lambda) + i\phi(\sigma\lambda)$.
Forming the ratio
\begin{eqnarray}
   R(\lambda, t) = { {\rm Im}C_+(\lambda,t) - {\rm Im}C_-(-\lambda,t)
                   \over
                   {\rm Re}C_+(\lambda,t) - {\rm Re}C_-(-\lambda,t) }
                 = - \tan(\phi(\lambda)t - \delta(\lambda))  \,,
\end{eqnarray}
with effective phase shift
\begin{eqnarray}
   \phi(\lambda) 
      = {1 \over t}\tan^{-1}\left(-R(\lambda,t)\right)\,,
         \qquad \mbox{where} \quad
          \phi(\lambda) = \phi_0\lambda + \phi_1\lambda^3
              + \ldots \,,
\end{eqnarray}
giving
\begin{eqnarray}
   \Delta q_B
      = \left. {\partial \phi (\lambda) \over 
                      \partial \lambda } \right|_{\lambda=0} \,.
\end{eqnarray}
This expression also holds for the connected piece and a test has
been performed for the connected piece using the imaginary signal,
to demonstrate its feasibility. (But of course it is better to use
in this case the form where no imaginary piece develops.)


\section{$SU(3)$ flavour symmetry breaking quark mass expansion}


In \cite{Bietenholz:2011qq} we developed $SU(3)$ flavour breaking
expansions for hadron masses for $2+1$ flavours and extended it to
matrix elements in \cite{Cooke:2012xv}. We follow and extend the
results given there (here just to `leading order' or LO). The flavour
structure is given from
\begin{eqnarray}
   A^I      = {1\over\sqrt{3}}
              \left( \bar{u}\gamma u + \bar{d}\gamma d + \bar{s}\gamma s \right)
                   \,, \quad
   A^{\pi^0} = {1\over\sqrt{2}}
             \left( \bar{u}\gamma u - \bar{d}\gamma d \right)
                   \,, \quad
   A^\eta   = {1\over\sqrt{6}}
             \left( \bar{u}\gamma u + \bar{d}\gamma d - 2\bar{s}\gamma s \right)
\end{eqnarray}   
where $\gamma \sim \gamma_i\gamma_5$, $A$ for the axial current.
So we can solve for $\bar{q}\gamma q \sim \Delta q$ in terms of
$A^I$, $A^{\pi^0}$ and $A^\eta$.
$SU(3)$ flavour breaking expansions for $A^{\pi^0}$, $A^\eta$,
are given in \cite{Cooke:2012xv}. In addition for the singlet operators,
$A^I$ we need to consider $8\times 1\times 8$ tensors, which are
similar to the mass expansions, \cite{Bietenholz:2011qq}.

We now consider the quark line `connected' and `disconnected' pieces
separately and just give here the results for the disconnected part.
(Complete expansions will be given in \cite{QCDSF}.)
To LO we have for the $SU(3)$ flavour breaking
expansion for $A^I$ for the baryon octet
\begin{eqnarray}
   {1 \over \sqrt{3}} \Delta\Sigma_N^{\ind{dis}}
               &=& a_0^{\ind{dis}} +3a_1^{\ind{dis}}\delta m_l
                                                        \nonumber \\
   {1 \over \sqrt{3}} \Delta\Sigma_\Sigma^{\ind{dis}}
               &=& a_0^{\ind{dis}} -3a_2^{\ind{dis}}\delta m_l
                                                        \nonumber \\
   {1 \over \sqrt{3}} \Delta\Sigma_\Xi^{\ind{dis}}
               &=& a_0^{\ind{dis}} -3(a_1^{\ind{dis}}-a_2^{\ind{dis}})\delta m_l \,,
\label{DSig_expan}
\end{eqnarray}
together with $\Delta\Sigma_\Lambda^{\ind{dis}}/\sqrt{3}
= a_0^{\ind{dis}} +3a_2^{\ind{dis}}\delta m_l$
and $\Delta\Sigma_{N_s}^{\ind{dis}}/\sqrt{3}
= a_0^{\ind{dis}} -6a_1^{\ind{dis}}\delta m_l$,
where $\Delta\Sigma_B^{\ind{dis}} = \Delta u_B^{\ind{dis}} + \Delta d_B^{\ind{dis}}
+ \Delta s_B^{\ind{dis}}$. All the expansions used here are for
$2+1$ quark flavours, $m_u=m_d\equiv m_l$, $m_s$ and the `distance'
from the flavour symmetric point ($m_l = m_s$) is given by
$\delta m_l = m_l - \overline{m}$, \cite{Bietenholz:2011qq},
where $\overline{m}$ is the average quark mass, held constant in
simulations, so the expansion parameters remain constant.
We have extended the nucleon octet to include a fictitious nucleon
consisting of strange quarks, denoted by $N_s$. (As well as the
$N$, $\Sigma$ and $\Xi$ this state can also be measured in a lattice
simulation.) As we are primarily interested in the nucleon, and
hence just $a_1^{\ind{dis}}$, it is convenient to consider the average of
the $\Sigma$ and $\Xi$ expansions
\begin{eqnarray}
   {1 \over 2\sqrt{3}} (\Delta\Sigma_\Sigma^{\ind{dis}}
      + \Delta\Sigma_\Xi^{\ind{dis}} )
         = a_0^{\ind{dis}} - {3\over 2} a_1^{\ind{dis}} \delta m_l \,.
\label{av_Sig_Xi}   
\end{eqnarray}
Using the above results together with those for $A^{\pi^0}$
and $A^\eta$ gives the separate expansions of
\begin{eqnarray}
   \Delta s^{\ind{dis}}_N
      &=& {1\over\sqrt{3}}a_0^{\ind{dis}}
           + \left( \sqrt{3}a_1^{\ind{dis}} -{2\over\sqrt{6}}r_1^{\ind{dis}} \right)
                  \delta m_l
                                                              \nonumber \\
   \Delta s^{\ind{dis}}_\Sigma
      &=& {1\over\sqrt{3}}a_0^{\ind{dis}}
          + \left( -\sqrt{3}a_2^{\ind{dis}} -{2\over\sqrt{6}}r_1^{\ind{dis}} \right)
                  \delta m_l
                                                              \nonumber \\
   \Delta s^{\ind{dis}}_\Xi
      &=& {1\over\sqrt{3}}a_0^{\ind{dis}}
             + \left( -\sqrt{3}(a_1^{\ind{dis}}-a_2^{\ind{dis}})
                    -{2\over\sqrt{6}}r_1^{\ind{dis}} \right)
                  \delta m_l \,,
\end{eqnarray}
together with
$\Delta s^{\ind{dis}}_\Lambda = a_0^{\ind{dis}}/\sqrt{3}
+ ( \sqrt{3}a_2^{\ind{dis}}
- 2(r_1^{\ind{dis}}+2r_2^{\ind{dis}})/\sqrt{6})\delta m_l$,
$\Delta s^{\ind{dis}}_{N_s} = a_0^{\ind{dis}}/\sqrt{3}
+ ( -2\sqrt{3}a_2^{\ind{dis}}-2r_1^{\ind{dis}}/\sqrt{6})\delta m_l$. 
Due to isospin invariance we have
$\Delta q_p^{\ind{dis}} = \Delta q_n^{\ind{dis}} \equiv \Delta q_N^{\ind{dis}}$,
$\Delta q_{\Sigma^+}^{\ind{dis}}
  = \Delta q_{\Sigma^-}^{\ind{dis}} \equiv \Delta q_{\Sigma}^{\ind{dis}}$,
$\Delta q_{\Xi^0}^{\ind{dis}}
  = \Delta q_{\Xi^-}^{\ind{dis}} \equiv \Delta q_\Xi^{\ind{dis}}$,
for $q = u$, $d$, $s$. Note that due to constraints, the cancellation of the
disconnected piece in $\Delta u_N^{\ind{dis}} - \Delta d_N^{\ind{dis}}$,
$\Delta u_\Sigma^{\ind{dis}} - \Delta d_\Sigma^{\ind{dis}}$ and
$\Delta u_\Xi^{\ind{dis}} - \Delta d_\Xi^{\ind{dis}}$
leads to the vanishing of $f^{\ind{dis}}$, $d^{\ind{dis}}$, $r_3^{\ind{dis}}$,
$s_1^{\ind{dis}}$, $s_2^{\ind{dis}}$ in \cite{Cooke:2012xv}. The results
are more complicated for the `connected' pieces; there are less
constraints, \cite{QCDSF}.

Useful results are here to consider a `singlet of singlets'
and the strange quark terms alone
\begin{eqnarray}
   X_{\Delta\Sigma}^{\ind{dis}}
      \equiv {1 \over 3} ( \Delta\Sigma_N^{\ind{dis}}
                      + \Delta\Sigma_\Sigma^{\ind{dis}}
                      + \Delta\Sigma_\Xi^{\ind{dis}} )
      &=& \sqrt{3} a_0^{\ind{dis}}
                                                          \nonumber \\
   {1 \over 3}\left( \Delta s_N^{\ind{dis}} + \Delta s_\Sigma^{\ind{dis}}
                   + \Delta s_\Xi^{\ind{dis}} \right)
       &=& {1 \over\sqrt{3}}a_0^{\ind{dis}}
                - \sqrt{2\over 3}r_1^{\ind{dis}}\delta m_l \,,
\label{sep_det}                   
\end{eqnarray}
which together with eq.~(\ref{av_Sig_Xi}) allow separate determinations
of $a_0^{\ind{dis}}$, $a_1^{\ind{dis}}$ and $r_1^{\ind{dis}}$.


\section{Renormalisation}


As the axial non-singlet currents $A^{\pi^0}$ and $A^{\eta}$ are (partially)
conserved currents, they have no anomalous dimensions
and so are scheme and scale independent. However the singlet current,
$A^I$ is no longer conserved if $n_f \not= 0$, as a topological term
$\propto 2n_f (\alpha_s/4\pi) F_{\mu\nu} \widetilde{F}_{\mu\nu}$
appears in the Ward identity. Thus we expect the renormalisation constant to
become scheme and scale dependent.
It is also convenient to again consider the renormalisation of the quark
line connected and disconnected pieces separately. We find
\cite{QCDSF:2011aa,Green:2017keo,Liang:2018pis}
\begin{eqnarray}
   \Delta q^{\ind{con}\,R} 
      = Z_A \Delta q^{\ind{con}}
        \,, \qquad
   \Delta q^{\ind{dis}\,R} 
      = Z_A \Delta q^{\ind{dis}} +
            {1\over 3} (Z_A^{\rm S} - Z_A) 
             (\Delta \Sigma^{\ind{con}} + \Delta \Sigma^{\ind{dis}}) \,,
\label{q_condis_R}
\end{eqnarray}
where $Z_A$ is the non-singlet renormalisation and $Z_A^{\rm S}$
is the singlet renormalisation factor.
This gives
\begin{eqnarray}
   \Delta \Sigma^{\ind{con}\,\rm R}
      = Z_A \Delta \Sigma^{\ind{con}} \,, \qquad
   \Delta \Sigma^{\ind{dis}\,\rm R}
      = Z_A^{\rm S} \Delta \Sigma^{\ind{dis}}
           + (Z_A^{\rm S} - Z_A)\Delta \Sigma^{\ind{con}} \,.
\end{eqnarray}


\section{Results and Conclusions}


We have a pion mass range from the flavour symmetric
point $M_\pi \sim 460\,\mbox{MeV}$ down to $\sim 300\,\mbox{MeV}$
on $a \sim 0.074\,\mbox{fm}$, $32^3\times 64$ lattices and configurations
as given in Table~\ref{all_parms}.
\begin{table}[!h]
   \begin{center}
      \begin{tabular}{c|rr|rr}
         Data set $\#$ &
         \multicolumn{1}{c}{$\kappa_l$}  & \multicolumn{1}{c}{$\kappa_s$}  & 
         \multicolumn{1}{c}{$\lambda_l$} &  \multicolumn{1}{c}{$\lambda_s$} \\  
       \hline
         1 & \multicolumn{2}{c|}{0.120900} & \multicolumn{2}{c}{-0.00625}  \\
         2 & \multicolumn{2}{c|}{0.120900} & \multicolumn{2}{c}{-0.0125
                                                             \phantom{0}}  \\
         3 & \multicolumn{2}{c|}{0.120900} & \multicolumn{2}{c}{0.0300}    \\
       \hline
         4 & 0.121095 & 0.120512  & 0.0000  & 0.0500                       \\
         5 & 0.121095 & 0.120512  & \multicolumn{2}{c}{-0.0250}            \\
         6 & 0.121095 & 0.120512  & \multicolumn{2}{c}{-0.0750}            \\
      \end{tabular}
   \end{center}
   \caption{Data sets used in the analysis.}
\label{all_parms}
\end{table}

\begin{figure}[!h]
   \begin{center}
   \begin{minipage}{0.40\textwidth}
    \includegraphics[width=5.50cm]{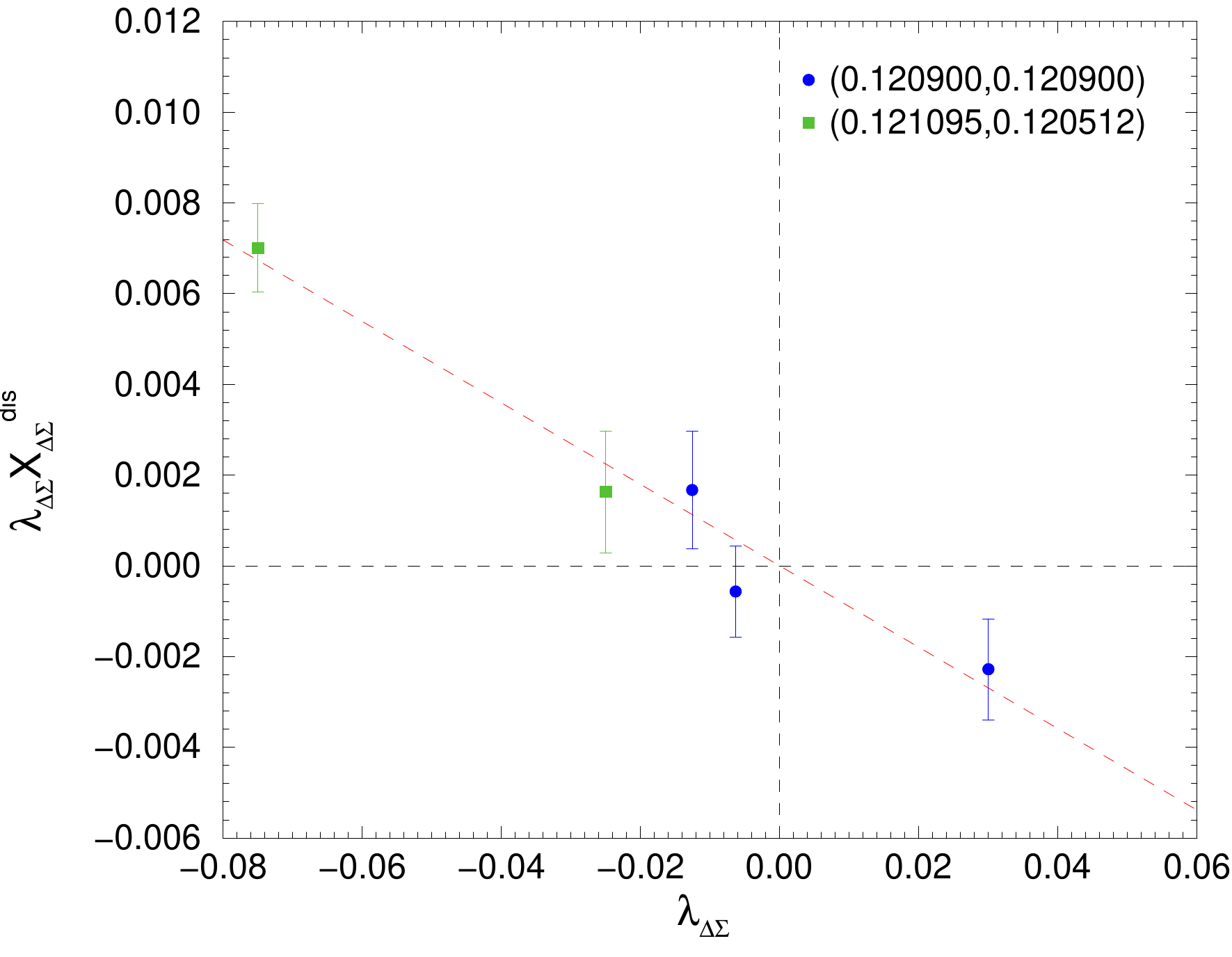}
   \end{minipage} \hspace*{0.10\textwidth}
   \begin{minipage}{0.40\textwidth}
      \vspace*{-0.15in}
      \includegraphics[width=5.00cm]{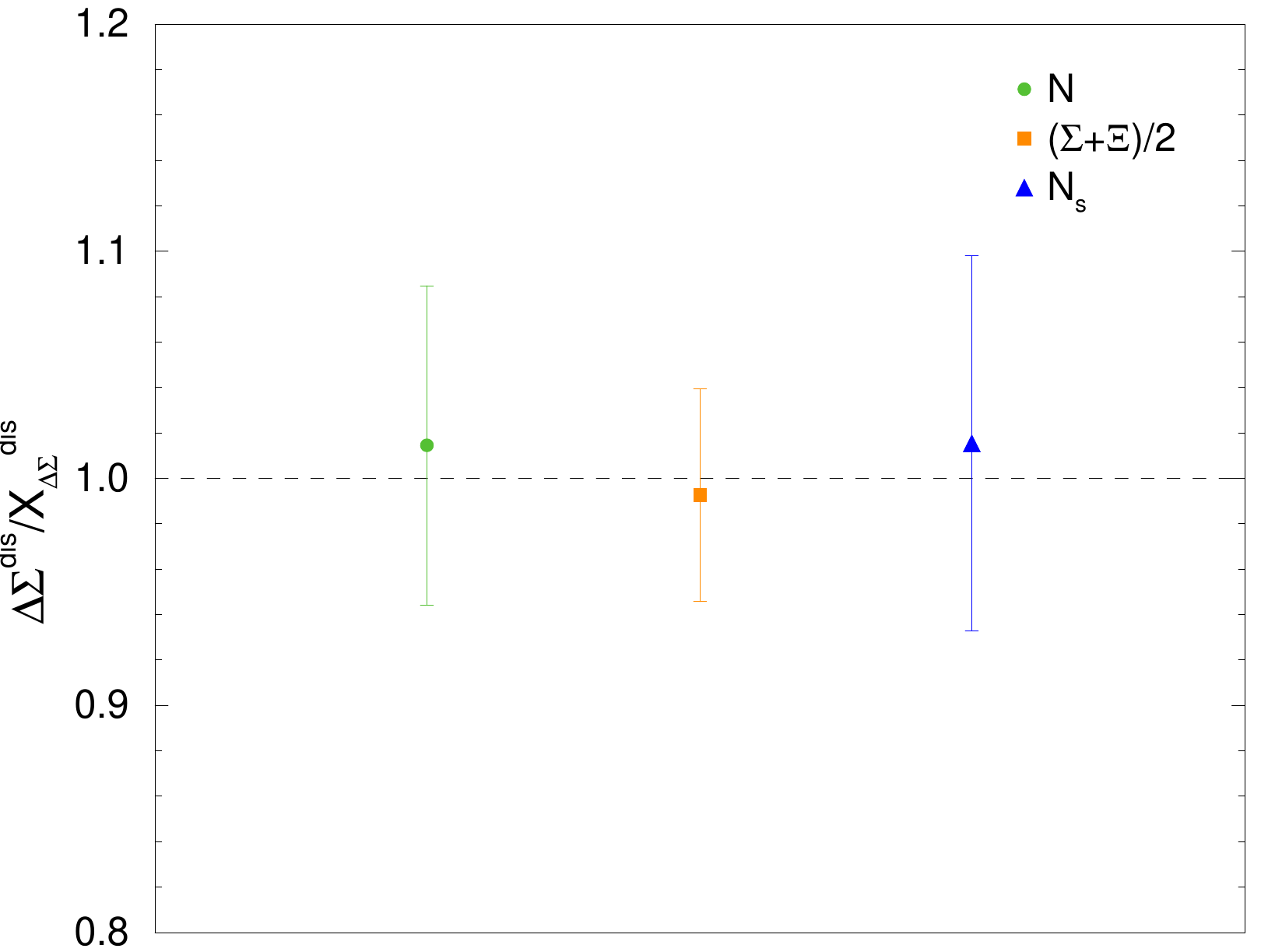}
   \end{minipage}
   \end{center}
   \vspace*{-0.15in}
   \caption{Left panel:
            $\phi_{\Delta\Sigma} = \lambda_{\Delta\Sigma}X_{\Delta\Sigma}^{\ind{dis}}$
            together with a linear fit using data sets $1$--$3$ and $5$, $6$;
            Right panel:
            $\Delta\Sigma_N^{\ind{dis}}/X_{\Delta\Sigma}^{\ind{dis}}$,
            $(\Delta\Sigma_\Sigma^{\ind{dis}}
               +\Delta\Sigma_\Xi^{\ind{dis}})/(2X_{\Delta\Sigma}^{\ind{dis}})$ and
            $\Delta\Sigma_{N_s}^{\ind{dis}}/X_{\Delta\Sigma}^{\ind{dis}}$
            for data set $6$.}
   \label{X_fan}
\end{figure}

We first consider $X_{\Delta\Sigma}^{\ind{dis}}$. In the left panel of
Fig.~\ref{X_fan} we show 
$\phi_{\Delta\Sigma} = \lambda_{\Delta\Sigma} X_{\Delta\Sigma}$, from eq.~(\ref{sep_det})
the gradient gives an estimation of $\sqrt{3}a_0^{\ind{dis}}$. Note that
we can now use all the available data sets, $1$--$6$, to determine
$X_{\Delta\Sigma}^{\ind{dis}}$ and hence $a_0^{\ind{dis}}$.

In the RH panel of Fig.~\ref{X_fan} we show
$\Delta\Sigma_N^{\ind{dis}}/X_{\Delta\Sigma}^{\ind{dis}}$,
$(\Delta\Sigma_\Sigma^{\ind{dis}}
               +\Delta\Sigma_\Xi^{\ind{dis}})/(2X_{\Delta\Sigma}^{\ind{dis}})$ and
$\Delta\Sigma_{N_s}^{\ind{dis}}/X_{\Delta\Sigma}^{\ind{dis}}$ for data set 6.
From eqs.~(\ref{DSig_expan},\ref{av_Sig_Xi}) we expect the
numerical values of $1 + 3a_1^{\ind{dis}}/a_0^{\ind{dis}}\delta m_l$,
$1 + 3/2a_1^{\ind{dis}}/a_0^{\ind{dis}}\delta m_l$ and
$1 - 6a_1^{\ind{dis}}/a_0^{\ind{dis}}\delta m_l$ (where $\delta m_l \sim -0.07$)
for $N$, $(\Sigma + \Xi)/2$ and $N_s$ respectively. We presently see
very little pattern in the data, so presently we take $a_1^{\ind{dis}} \approx 0$.
This indicates that this disconnected part is very small for all the
baryons in the octet.
A tentative general conclusion is that there is very little sign of $SU(3)$
flavour symmetry breaking effects in the disconnected pieces.
Furthermore with $a_1^{\ind{dis}} \approx 0$ this also implies that
\begin{eqnarray}
   \Delta s_N^{\ind{dis}}
     \approx
     {1 \over 3}\left( \Delta s_N^{\ind{dis}} + \Delta s_\Sigma^{\ind{dis}}
                       + \Delta s_\Xi^{\ind{dis}} \right) \,,
\end{eqnarray}
also away from the $SU(3)$ flavour symmetry point.
So when using data set 4 we can avoid a direct determination of
$r_1^{\ind{dis}}$.

\begin{figure}[!h]
   \begin{center}
      \includegraphics[width=5.50cm]{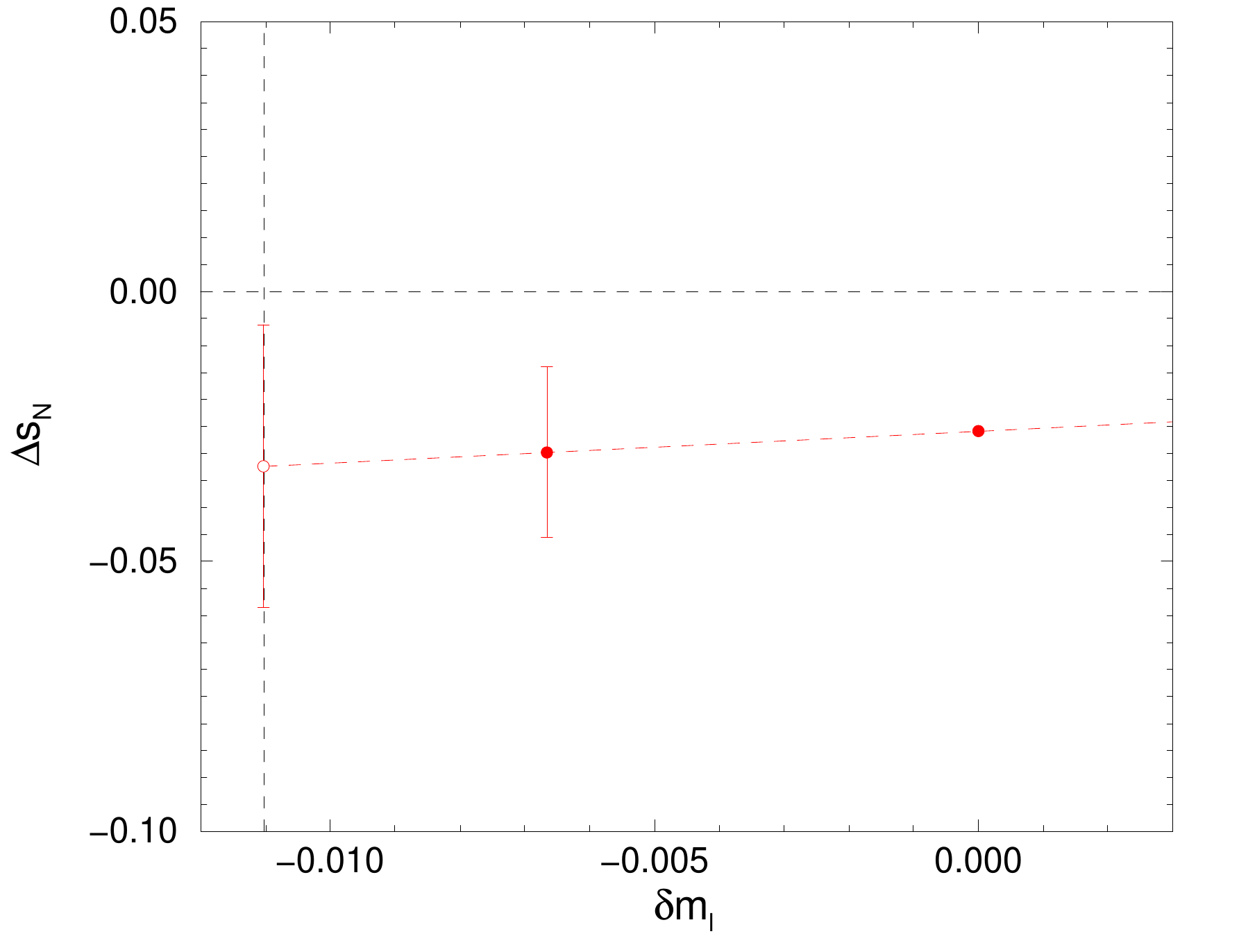}
   \end{center}
\vspace*{-0.15in}
\caption{$\Delta s_N$ in the $\overline{MS}$ scheme at a scale of
         $2\,\mbox{GeV}$ versus $\delta m_l$. The vertical line
         indicates where the physical pion mass lies,
         \protect\cite{Horsley:2014koa}.}
\label{deltasR}   
\end{figure}

We have computed $Z_A$ and $Z_A^{\rm S}$ at $2\,\mbox{GeV}$
in \cite{Chambers:2014pea}, also using the FH method to give
$Z_A = 0.8458(8)$, $Z_A^{\rm S}(2\,\mbox{GeV}) = 0.8662(34)$
(the latter in the $\overline{MS}$ scheme). Note that this means
that, as expected $(Z_A^{\rm S} - Z_A)/Z_A^{\rm S} \sim 2\%$
a small difference, which we shall presently ignore.
In Fig.~\ref{deltasR} we show the renormalised results for $\Delta s_N$
in the $\overline{MS}$ scheme at a scale of $2\,\mbox{GeV}$.
Linearly extrapolating to the physical pion mass we find a
preliminary result of $\Delta s_N(2\,\mbox{GeV}) = -0.032(26)$.

In conclusion `disconnected' quantities are notoriously difficult
quantities to compute
as they are a short distance quantity and suffers from large fluctuations.
As alternative to more standard `stochastic' approaches
we have developed a method using the Feynman--Hellmann theorem,
together with a $SU(3)$ flavour breaking expansion.


\section*{Acknowledgements}


The numerical configuration generation (using the BQCD lattice
QCD program) and data analysis (using the Chroma library)
was carried out
on the IBM BlueGene/Q and HP Tesseract using DIRAC 2
(EPCC, Edinburgh, UK), the IBM BlueGene/Q (NIC, J\"ulich, Germany)
the Cray XC40 (HLRN, The North-German Supercomputer
Alliance) and the NCI National Facility in Canberra, Australia
(supported by the Australian Commonwealth Government).
HP was supported by DFG Grant No. PE 2792/2-1.
PELR was supported in part by the STFC under contract ST/G00062X/1
and RDY and JMZ were supported by the Australian Research Council Grants
FT120100821, FT100100005 and DP140103067.
RH wishes to thank G. Shore for a useful discussion.



\end{document}